\documentclass[a4paper,11pt]{article}

\usepackage{letter}
\usepackage{colortbl}
\usepackage{soul}
\usepackage{xltxtra} 
\usepackage{xgreek} 
\usepackage{fontspec}
\usepackage{listings}
\usepackage{graphicx}
\usepackage{fancyhdr}

\usepackage[english]{babel}
\usepackage{float}

\usepackage{colortbl}
\usepackage[table]{xcolor}
\definecolor{lightgray}{gray}{0.9}

\usepackage[margin=2.5cm]{geometry}

\usepackage[vlined,linesnumbered]{algorithm2e}

\SetKwInOut{Input}{Input}
\SetKwInOut{Output}{Output}
\SetKwInOut{Parameter}{Parameter}
\SetKwInput{Algorithm}{Algorithm}
\SetKwInput{Remark}{Remark}
\SetKw{Let}{Let}
\SetKw{Add}{Add}
\SetKwComment{Comment}{{// }}{}

\usepackage[backend=biber,sorting=none]{biblatex}
\definecolor{customgreen}{rgb}{0,0.6,0}
\definecolor{customgray}{rgb}{0.5,0.5,0.5}
\definecolor{custommauve}{rgb}{0.6,0,0.8}
\lstnewenvironment{SQLcode}{%
\lstset{language=SQL,%
showspaces=false,%
showstringspaces=false,%
tabsize=4,%
lineskip=2pt%
}}
{}

\lstnewenvironment{latex}{%
\lstset{language=TeX,%
keywordstyle=\bf\sffamily\color{blue},%
commentstyle=\color{ForestGreen},%
stringstyle=\color{Maroon},%
basicstyle=\ttfamily\normalsize\color{black},%
showspaces=false,%
showstringspaces=false,%
tabsize=4,%
aboveskip=10pt,%
belowskip=10pt,%
lineskip=2pt%
}}
{}

\usepackage{float}
\usepackage{amsmath}
\title{
\huge Evaluating Future Air Traffic Management Security
}

\author{Konstantinos Spalas, MSc in Computer Science, Tripoli, dit2318cst@go.uop.gr \\
}

\date{11 June 2025}

\begin{document}

\maketitle


%
%
\begin{center}
  \section*{Abstract}  
\end{center}

The L-Band Digital Aviation Communication System (LDACS) aims to modernize communications between the aircraft and the tower. Besides digitizing this type of communication, the contributors also focus on protecting them against cyberattacks. There are several proposals regarding LDACS security, and a recent one suggests the use of physical unclonable functions (PUFs) for the authentication module. This work demonstrates this PUF-based authentication mechanism along with its potential vulnerabilities. Sophisticated models are able to predict PUFs, and, on the other hand, quantum computers are capable of threatening current cryptography, consisting factors that jeopardize the authentication mechanism giving the ability to perform impersonation attacks. In addition, aging is a characteristic that affects the stability of PUFs, which may cause instability issues, rendering the system unavailable. In this context, this work proposes the well-established Public Key Infrastructure (PKI), as an alternative solution.
\begin{center}
\[\]
\textbf{Key words} : PUF, LDACS, Aviation, Communications, Post-Quantum
\end{center}
%
%
\section{Introduction}\label{sec:intro}
By now, communications between the aircraft and the control tower utilize analog RF signals. However, the growing congestion of radio frequencies, especially at popular airports, implies the need to upgrade the communication infrastructure, mitigating the risks when multiple aircraft request tower radio contact. In addition, current communication systems lack resilience against cyberattacks \cite{no-sec}. Hence, the Single European Sky ATM\footnote{Air Traffic Management} Research (SESAR) \cite{sesar} takes on this challenge, making ATM modern by addressing a number of systematic problems. One of its pillars is to establish the L-Band Digital Aviation Communication System (LDACS) \cite{ldacs}, \cite{ldacs-2}, which will serve our skies in the future, promising to include Post-Quantum Cryptography (PQC). 

The main purpose of this work is to evaluate a recent PUF-based authentication mechanism of LDACS \cite{pmake}, \cite{ldacs-puf}. In return, a PQC- based PKI is proposed as a countermeasure. The following Ch. \ref{background} briefly refers to the background in order to support the evaluation developed in Ch. \ref{protocol}.
%
%
\section {Background}\label{background}
\textit{PUF}: A Physical Unclonable Function (PUF) \cite{PUF} is a hardware-based security primitive that leverages the unique physical characteristics of a device to produce a distinct response $R$ when given a specific input, named challenge $C$.  PUFs are commonly used for secure authentication, key generation, and protection against hardware cloning.

\textit{ICAO address}: The International Civil Aviation Organization address $\text{ICAO}_\text{A}$\footnote{https://en.wikipedia.org/wiki/ICAO\_code} is a unique 24-bit identifier assigned to every aircraft transponder. Issued by national aviation authorities, this address ensures global aircraft identification and tracking, enabling reliable communication between aircraft and ground-based surveillance systems.

\textit{CMA-ES ML algorithm}\label{sec:cma-es}:
Covariance Matrix Adaptation Evolution Strategy (CMA-ES) is an advanced evolutionary algorithm designed for continuous optimization problems. The authors in \cite{CMA-ES} used the CMA-ES algorithm to predict the behavior of PUFs with great success. Ideally, CMA-ES is best used to optimize Machine Learning attributes.

\textit{Quantum computers}: Advanced machines that use the principles of quantum mechanics to process information. Unlike classical computers, which use bits (0s and 1s) to represent data, quantum computers use quantum bits (qubits). Lov Grover \cite{grover} has proven that quantum computers can reduce the cost of a search to $\mathcal{O}(2^{n/2})$, rather than the classical computer $\mathcal{O}(2^{n})$.
%
%
\section{The authentication protocol}\label{protocol}
To understand the aforementioned lightweight PUF-based authentication mechanism for LDACS, first we must go through the \textit{registration} phase, which is a one-time procedure, performed offline prior to system deployment. It ensures that the aircraft and tower secretly share the cryptographic attributes necessary for mutual authentication in the open air. Thus, the former's radio unit stores $\tau$ and $\theta=h(R)$, while the latter's stores $\text{ICAO}_\text{A}$, $C$, $R$, $\ \tau$, where:  
\begin{equation}\label{eq:t}
    \tau = h (\text{ICAO}_\text{A} || R)_{24}
\end{equation}
\begin{equation}\label{eq:r}
    R = PUF(C)
\end{equation}
with $h$ to be a hash function. Note that $\tau$ contains the first 24 bits of the hash.

\begin{figure}
\centering
\includegraphics[scale=.4]{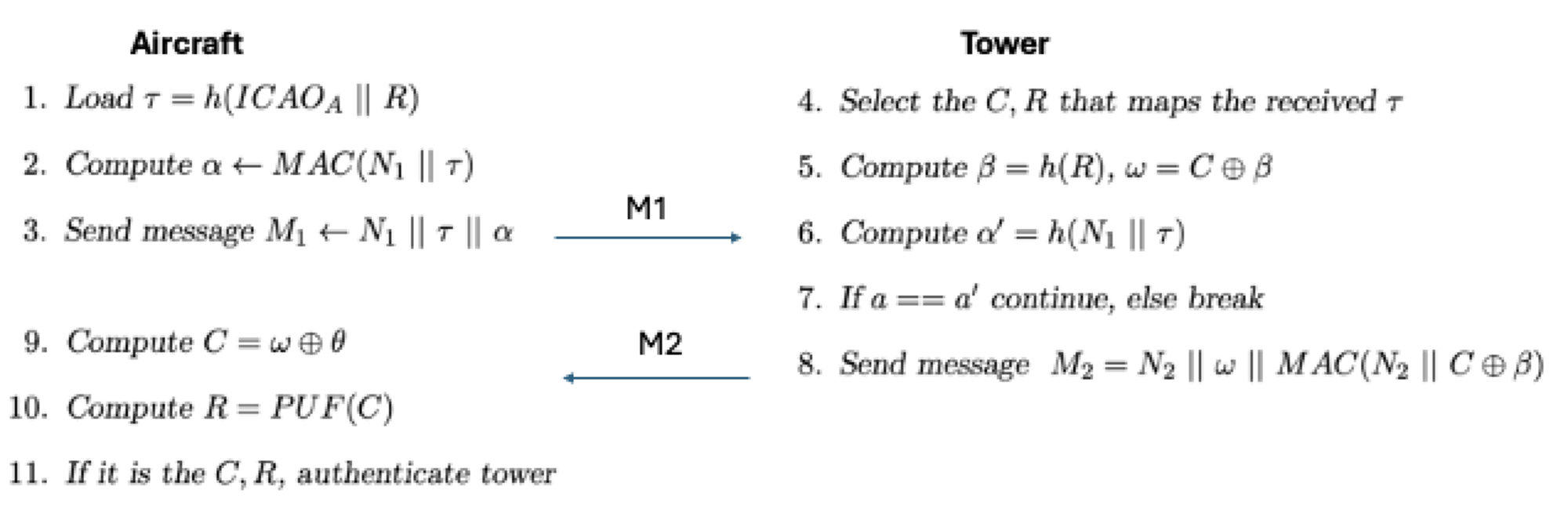}
\caption{Aircraft authenticates tower}
\label{fig:auth}
\end{figure}

After being registered, a radio unit can serve an aircraft. So, if pilots request radio contact during flight or even before engines start, the authentication protocol exchanges several messages. The process leads the system to produce the same symmetric key for a secure channel, utilizing Key Encapsulation Mechanism (KEM). This operation is illustrated in Fig. \ref{fig:auth}, until the step where the aircraft authenticates the tower. MAC is denoted as a Message Authentication Code and $N_1$ and $N_2$ are nonce values to keep communication fresh.

%
%
\subsection{Leakage of $C$, $R$ }
Despite that crucial information sent is hashed, someone could receive these signals using the appropriate hardware \cite{cpdl}, isolating any part of the messages for further processing. So, he can easily extract the value $\tau$ from the message $M_1$ because he knows its structure \cite{ldacs-puf}. Consequently, in a small airport, where flights are rare, it would be easy to match $\tau$ with the corresponding $\text{ICAO}_\text{A}$, which is a public and permanent value that uniquely describes an aircraft.  For the tower point of view, the pseudo-address $\tau$ functions as a filter on the $C$, $R$ pair look-up table, as depicted in Fig. \ref{fig:auth} and thus it might also be unique for each aircraft.

Moreover, in \cite{puf-lenth} is mentioned that an ideal PUF would consist of a 32-bit challenge $C$ and a 128-bit response $R$. As these deterministic functions are one-to-one, the challenge set $\mathcal{C} = \{0,1\}^{32}$ maps $2^{32}$ responses to the set $\mathcal{R} = \{0,1\}^{128}$. Consequently, the $C$, $R$  pair that interests an adversary is the one that solves both Eq. \ref{eq:t}, \ref{eq:r}.

 \textit{Attack, method 1}: Recall Ch. \ref{sec:cma-es} where the authors managed to predict the behavior of several PUFs utilizing the sophisticated CMA-ES ML technique. In this context, if an attacker cannot precisely replicate a PUF chip using advanced fabrication techniques or cannot acquire one produced by the same vendor, a well-trained model might be able to predict the behavior of the embedded PUF. In this case, the model will map $\mathcal{C} \rightarrow \mathcal{R}$, and then Eq. \ref{eq:t}, \ref{eq:r} will be used by the adversary to create a look up table with the values $\tau$, $\textit{ICAO}_\textit{A}$, $C$ and $R$. The Alg. \ref{Alg} presents the rationale for disclosing the secret attributes of a PUF-based ATM. Note that the aforementioned CMA-ES algorithm can mitigate the error rate, predicting $R$, to $98\%$. Hence, for 128 and 192 bit $R$, the error $2\%$ is equal to 2 and 3 bit flips (BF), respectively. So, to predict it with precision, we must further calculate the permutations $P(|R|,BF)$. Then $P(128,2) \approx 16\cdot10^3$ and $P(192,3) \approx 7\cdot10^6$, which are feasible using classical computing. In this kind of attack, the cost does not increase exponentially with respect to the length of $R$.

\vspace{2mm}
\begin{algorithm}[H]
\label{Alg}
\caption{Attack, method 1}
\KwIn{Many PUFs, $M_1$, $\tau$ }
\KwOut{A matching pair $(C, R)$ such that $h(ICAO_{\textit{A}} || R) = \tau$}
Map each $ICAO_{\textit{A}}$ to corresponding $\tau$\;
Success $\gets$ \textbf{false}\;
\Repeat{Success}{
    Train model to map vectors from set $\mathcal{C} \rightarrow \mathcal{R} $\;
    Select a $\tau$\;
    \ForEach{$C \in \mathcal{C}$}{
        Select corresponding $R$ from $\mathcal{R}$\;
        Compute $h(ICAO_{\textit{A}} || R)$\;
        \If{$\tau = h(ICAO_{\textit{A}} || R)$}{
            Success $\gets$ \textbf{true}\;
            \Return{$(C, R)$}\;
        }
    }
}
\end{algorithm}
\vspace{2mm}
     \textit{Attack, method 2}: Both $\textit{ICAO}_\textit{A}$ and $\tau$ have 24 bit lengths, and each of them uniquely identifies an aircraft, the former as its real address, whereas the latter as its pseudo-address. Going through Eq. \ref{eq:t}, Fig. \ref{fig:auth} and the protocol in Ch. \ref{protocol}, we can deduce that a given $\textit{ICAO}_\textit{A}$ corresponds to certain $R$ and $\tau$. Hence, for a single $\tau$, thereby a single aircraft, the corresponding $R$ is hidden in the set $\mathcal{R} = \{0,1\}^{128}$ and is the one that solves Eq. \ref{eq:t}. In such a case, the quantum preimage algorithm \cite{preimage} can reduce the total cost of a brute-force search from $\mathcal{O}(2^{n})$ to $\mathcal{O}(2^{n/3}) = \mathcal{O}(2^{128/3}) \approx \mathcal{O}(2^{42}$), which is feasible in negotiable time using quantum computers\footnote{https://www.ibm.com/roadmaps/quantum/2030/} future  capabilities. Finally, the challenge $C$ can be revealed by solving $\omega = C \oplus h(R)$, where $\omega$ is a known value transmitted from the tower to the aircraft and $h$ a hash function.

%
%
\subsection{PUF aging}
Aging is inevitable in PUFs, but there are sometimes ways to mitigate it. In any case, \cite{aging} mentions that aging increases by $19\%$ every two years, and in this context, consideration arises about their reliability on authentication mechanisms.
%
%
\section{Conclusions and proposals}
While authorities tend to transform future aviation communications to a modern and more secure environment with PUF-based authentication mechanisms being an option, the strength of future quantum computation, along with the ML algorithm, threatens the security resilience. This results in the need for a more secure and latency-free resistance system. Instead, a well-established PQC-based PKI between the aircraft, tower, and a certification authority could be a significant alternative.

\printbibliography

\end{document}